\documentclass[
 article,
 twocolumn,
 groupedaddress,
 showpacs,
 preprintnumbers,
 longbibliography,
 amsmath,
 amsthm,
 amssymb,
 aps,
 prx,
 floatfix,
]{revtex4-1}

\usepackage{graphicx}					
\usepackage{verbatim}					
\usepackage{dcolumn}                    
\usepackage[many]{tcolorbox}
\usepackage[ruled,vlined]{algorithm2e}
\usepackage{color}
\usepackage{fontawesome}
\usepackage{adjustbox}

\usepackage{orcidlink}                     

\newtheorem{theorem}{Theorem}

\newcommand{\ds}{\displaystyle}             

\newcommand{\dd}{\mathrm{d}}                

\newcommand{\Rs}{\mathcal{R}}                 

\newcommand{\ob}{\mathcal{O}}                

\newcommand{\K}{\mathcal{K}}                
\newcommand{\cs}{\mathrm{C.S.}}             

\newcommand{\z}{\zeta}                      

\begin{document}

\title{
Geometry of Almost-Conserved Quantities in Symplectic Maps
\\ Part II. Recovery of approximate invariant
}
\author{T.~Zolkin\,\orcidlink{0000-0002-2274-396X}}
\email{iguanodyn@gmail.com}
\affiliation{Independent Researcher, Chicago, IL}
\author{S.~Nagaitsev\,\orcidlink{0000-0001-6088-4854}}
\affiliation{Brookhaven National Laboratory, Upton, NY 11973}
\affiliation{Old Dominion University, Norfolk, VA 23529}
\author{I.~Morozov\,\orcidlink{0000-0002-1821-7051}}
\affiliation{Synchrotron Radiation Facility ``SKIF'', Koltsovo 630559, Russia}
\affiliation{Novosibirsk State Technical University, Novosibirsk 630073, Russia}
\author{S.~Kladov\,\orcidlink{0000-0002-8005-9373}}
\affiliation{University of Chicago, Chicago, IL 60637}

\date{\today}

\begin{abstract}
Noether's theorem, which connects continuous symmetries to exact
conservation laws, remains one of the most fundamental principles
in physics and dynamical systems.
In this work, we draw a conceptual parallel between two paradigms:
the emergence of exact invariants from continuous symmetries, and
the appearance of approximate invariants from discrete symmetries
associated with reversibility in symplectic maps.
We demonstrate that by constructing approximating functions that
preserve these discrete symmetries order by order, one can systematically uncover hidden structures, closely echoing Noether's
framework.
The resulting functions serve not only as diagnostic tools but also
as compact representations of near-integrable behavior.

The second article applies the method to global dynamics, with a
focus on large-amplitude motion and chaotic systems.
We demonstrate that the approximate invariants, once averaged,
accurately capture the structure of resonances and the boundaries
of stability regions.
We also explore the recovery of exact invariants in integrable
cases, showing that the method reproduces the correct behavior
when such structure is present.
A single unified function, derived from the map coefficients, yields
phase portraits, rotation numbers, and tune footprints that closely
match numerical tracking across wide parameter ranges.
Comparisons with the Square Matrix method reveal that while both
approaches satisfy local constraints, our technique provides greater
accuracy and robustness in resonant and strongly nonlinear regimes.
These results highlight the method's practical power and broad
relevance, offering a compact, analytic framework for organizing
nonlinear dynamics in symplectic maps with direct applications to
beam physics and beyond.

\end{abstract}

\maketitle

\section{Introduction}

\vspace{-0.2cm}
The first part of this study developed a systematic perturbative
method for constructing approximate invariants in symplectic maps
--- discrete systems closely related to Hamiltonian dynamics and
widely applicable across mathematical and physical problems
\cite{arnold1989ergodic,arnold1997celestial,arnold1988}.
These invariants are not only conserved up to a given order but
also respect two families of discrete symmetries induced by
reversibility~\cite{roberts1992revers}.
In continuous-time dynamics, Noether's theorem~\cite{Noether1918}
establishes a correspondence between symmetries of the action and
exact integrals of motion;
here, we uncover a similar correspondence in the discrete setting.

By benchmarking our method against results from Lie algebra
technique and by probing the structure of low-order resonances,
we demonstrated both the mathematical consistency and practical
accuracy of the approach in the vicinity of a fixed point.
In this second part, we shift the focus from foundational
formulation to broader applications.
Much like the classical Arnold scheme in KAM theory
\cite{kolmogorov1954conservation,arnol1963small,moser1962invariant},
as the order of approximation increases, the simply connected
domain where the approximate invariant holds shrinks toward a
point.
In this limit, the system approaches an integrable twist map ---
the same map that lies at the heart of Lie algebra benchmarking.
However, our interest lies beyond this infinitesimal regime:
we aim to extend the method into the realm of large-amplitude
dynamics.

On one end, we investigate near-integrable systems exhibiting
chaos beyond the fixed point neighborhood.
In such cases, the method uncovers coherent structures --- such
as chains of islands --- that are otherwise obscured by standard
twist expansions of action-angle variables.
On the other end, we revisit known integrable maps and test whether
the method can recover their exact global invariants.
This is a critical test: failure in the integrable case would cast
doubt on capturing global features in chaotic systems,
where no exact invariant exists.

\newpage
We begin where the first part left off, by briefly restating the
method for general planar maps.
The remainder of this article is organized as follows.
In Section~\ref{sec:Recovery}, we test the method on all known
integrable maps with smooth invariants for transformation in
McMillan form~\cite{mcmillan1971problem}, systematically
described by Suris~\cite{suris1989integrable}.
For maps with finite polynomial invariants, our results fully
recover the expected structure;
for maps whose invariants are infinite power series, we observe
almost-everywhere convergence.
In Section~\ref{sec:Chaos}, we examine the method's limitations
by applying it to chaotic systems, focusing on the quadratic and
cubic H\'enon maps~\cite{henon1969numerical}.
They are relatively well understood
\cite{sterling1999homoclinic,dullin2000twistless,dulin2000henon},
yet exhibit rich nonlinear dynamics, making them ideal for
benchmarking new methods against known results.
These maps are directly relevant to accelerator physics, modeling
horizontal dynamics in lattices with thin sextupole and octupole
lenses~\cite{zolkin2024MCdynamics,zolkin2024MCdynamicsIII}.
We use several case studies where the phase space contains large
island chains, analyzing both the approximate invariants and the
nonlinear rotation number derived from them, providing an
alternative to traditional twist expansion.
In Subsection~\ref{sec:Simply}, we extend this analysis to the
joint space of dynamical variables and map parameters, constructing
approximations of the simply connected domain of validity.
Building on recent work [Fractal], where integrable systems were
shown to approximate the geometry of resulting fractal structures
to first nonlinear order, we demonstrate that the perturbative
approach enables further refinement --- albeit with limitations
--- beyond the fixed points/$n$-cycles framework.

One crucial component of this method is the averaging procedure,
which resolves under-determinacy in the expansion coefficients
and ensures resonance compatibility
(i.e., applicability at rational rotation numbers).
As a case study, we compare this to the naive alternative of
setting all undetermined coefficients to zero.
Along the way, we draw comparisons with other analytical tools,
particularly the Square Matrix (SM) method, and show how even
low-order approximations can capture essential features of
strongly nonlinear systems.

In the following and final part of this series, we demonstrate
its effectiveness by applying it to realistic dynamical models
corresponding to accelerator lattice configurations from the
FermiLab complex.
As we show, this construction naturally extends the classic
Courant-Snyder~\cite{courant1958theory} formalism --- a foundational
framework in accelerator physics now spanning seven decades and
historically focused on linear phenomena.
This limitation stemmed from several factors: weak nonlinearities
at small amplitudes, design goals favoring linearity, and the
limited analytical tools available at the time (1953),
preceding the full development of KAM theory and later progress
in chaotic dynamics.
While this is not the first attempt to describe near-integrable
motion using normal forms, we believe the present approach holds
promise for broader applications due to its conceptual simplicity.

\section{Perturbation method}

Consider a map of the plane $\mathrm{T}$:
$\mathbb{R}^2\mapsto\mathbb{R}^2$, which is symplectic
(i.e., it preserves area and orientation) and admits a
power series expansion:
\[
\begin{array}{cc}
\ds q' = A_{1,0}\,q + A_{0,1}\,p + A_{2,0}\,q^2 + A_{1,1}\,q\,p +
A_{0,2}\,p^2 + \ldots,\\[0.25cm] 
\ds p' = B_{1,0}\,q + B_{0,1}\,p + B_{2,0}\,q^2 + B_{1,1}\,q\,p +
B_{0,2}\,p^2 + \ldots.
\end{array}
\]

\noindent$\bullet$
We begin by introducing a general polynomial ansatz for the
approximate invariant of order $n$:
\[
\K^{(n)}[p,q] = \K_0 + \epsilon\,\K_1 + \ldots + \epsilon^n\,\K_n,
\]
where $\K_m$ is a homogeneous polynomial of degree $(m+2)$:
\[
\K_m[p,q] = \sum_{\substack{i,j \geq 0 \\ i + j = m+2}}
C_{i,j}\,p^i q^j.
\]
To determine the coefficients $C_{i,j}$, we require that the
approximate invariant be conserved up to $\ob(\epsilon^{n+1})$:
\[
\Rs_n = \K^{(n)}[p',q'] - \K^{(n)}[p,q] = 
    \overline{\Rs_n}\,\epsilon^{n+1} +
    \ob(\epsilon^{n+2}).
\]
The lowest-order contribution, which can also be derived via
standard linearization, is defined only up to a constant
multiplier
\begin{equation}
\label{math:K0}    
\K_0[p,q]=C_0\left[
    A_{0,1}\,p^2+(A_{1,0}-B_{0,1})\,p\,q-B_{1,0}\,q^2
\right].
\end{equation}
The {\it seed} coefficient $C_0$ can be set to unity or chosen
specifically to eliminate resonant denominators.

\noindent$\bullet$
Due to under-determinacy, the invariant (after all two-indexed
coefficients $C_{i,j}$ have been fixed) is known only up to a
series:
\[
    C_0\,\K_0 + C_1\,\K_0^2 + \ldots.
\]
Nevertheless, the approximate invariance condition holds to the
required order, and the twist coefficient $\tau_k$ converges at
order $n = 2\,k + 2$.

\noindent$\bullet$
The remaining coefficients $C_k$ (for $k>0$) are obtained via the
{\it averaging procedure}, which minimizes the residual error.
After transforming to the eigenbasis of the Jacobian,
$(q,p)\rightarrow(Q,P)$, the linear part of the map becomes a pure
rotation, and the zeroth-order invariant takes the form
\[
\K_0[P,Q] = P^2 + Q^2.
\]
The coefficients $C_k$ are then determined by solving another
system of linear equations
\[
    \frac{\dd}{\dd C_k}\,I_n = 0,
\]
where the integral
\[
I_n = \int_0^{2\pi} \overline{\Rs_n}^2[\rho,\psi] \,\dd\psi.
\]
Here, $(\rho,\psi)$ are polar phase space coordinates defined by
$Q = \rho\cos\psi$ and $P = \rho\sin\psi$.

\newpage
\section{\label{sec:Recovery}Recovery of integrable systems}

In the first part of this manuscript, we explored various examples
demonstrating that knowledge of the twist coefficients alone is
insufficient to reconstruct the full invariant of motion, as the
coefficients $C_k$ remain undetermined.
Before venturing into applications for chaotic systems, it is
essential to validate our perturbative approach and the averaging
procedure on systems that possess an exact invariant of motion.
Since the PT was specifically designed to approximate such
invariants, integrable systems provide an ideal testing ground.

Fortunately, the McMillan form of the map, explored in the first
part of this manuscript, is not only one of the simplest forms of
a planar transformation, offering significant flexibility in its
dynamics through various choices of force functions, but it is also inherently symplectic and reversible, exhibiting clear symmetries
for any force function $f(p)$, not necessarily analytical.
Furthermore, for this form of the map, Suris' theorem
\cite{suris1989integrable} provides significant constraints on
the possible forms of the invariant when it is assumed to be an
analytic function.

\begin{theorem}[Yu. B. Suris]
Consider the McMillan form mapping $(q,p)\mapsto(p,-q+f(p))$ over
the field of complex numbers $\zeta=(q,p)\in\mathbb{C}^2$, where
$f(p)$ is holomorphic in some band of finite width (independent of $\epsilon$) about the real axis.
Then it has a nontrivial symmetric integral $\K[p,q]$, holomorphic
in the domain $|p-q| < \delta_0$, in the following and only in the
following three cases:
\begin{equation*}
\begin{array}{ll}
(\mathrm{I}):     &\ds  \mathcal{K}[p,q] =
    \mathrm{A}\,p^2q^2 + \mathrm{B}\,(p^2q + p\,q^2)            \\[0.25cm]
    & \ds\qquad\qquad + \,\,
        \Gamma\,(p^2 + q^2) + \mathrm{E}\,p\,q + \Delta\,(p+q), \\[0.35cm]
(\mathrm{II}):    &\ds  \mathcal{K}[p,q] =
    \mathrm{A}\,e^{\alpha\,p}e^{\alpha\,q} +
    \mathrm{B}\,(e^{\alpha\,p} + e^{\alpha\,q})                 \\[0.25cm]
    & \ds\qquad\qquad + \,\,
    \Gamma\,(e^{\alpha\,p}e^{-\alpha\,q} +
             e^{-\alpha\,p}e^{\alpha\,q})                       \\[0.25cm]
    & \ds\qquad\qquad + \,\,
    \Delta\,(e^{-\alpha\,p} + e^{-\alpha\,q}) +
    \mathrm{E}\,e^{-\alpha\,p}e^{-\alpha\,q},                   \\[0.30cm]
(\mathrm{III}):   &\ds \mathcal{K}[p,q] =
    \Lambda_1\,(\cos[\omega\,p-\psi]+\cos[\omega\,q-\psi])      \\[0.25cm]
    & \ds\qquad\qquad + \,\,
    \Lambda_2\,\cos[\omega\,(p+q)-\phi]                         \\[0.25cm]
    & \ds\qquad\qquad + \,\,
    \Lambda_3\,\cos[\omega\,(p-q)].
\end{array}
\end{equation*}
Here $\mathrm{A}$, $\mathrm{B}$, $\Gamma$, $\Delta$, $\mathrm{E}$,
$\Lambda_{1,2,3}$, $\alpha$, $\omega$, $\psi$ and $\phi$ are
arbitrary constants (guaranteeing the absence of singularities
for $f(p)$ in the neighborhood of the real axis).
\end{theorem}

\subsection{Symmetric McMillan map}

The first invariant corresponds to the {\it symmetric integrable
McMillan map}~\cite{mcmillan1971problem,IR2002II}, which has a
polynomial form that is biquadratic in $p$ and $q$.
This specific structure of the exact invariant naturally aligns
with the polynomial form of our approximate invariant.
For transformations with a fixed point at the origin ($\Delta=0$),
the invariant can, in almost all cases ($\Gamma \neq 0$), be
rewritten in a simplified form
\cite{zolkin2024MCdynamics,zolkin2024MCdynamicsIII}:
\begin{equation}
\label{math:K-I}    
\K_\mathrm{I}[p,q] = \cs + \beta\,\Pi\,\Sigma + \alpha\,\Pi^2.
\end{equation}
We adopt the symmetric notation $\Sigma = p + q$, $\Pi = p\,q$,
and define $\K_0[p,q] \equiv \cs = \Sigma^2 - r_2\,\Pi$.
The resonant denominators are denoted as $r_{1,2} = a \mp 2$,
$r_3 = a + 1$, and $r_4 = a$.
For full details, see the first part of the manuscript.
The corresponding force is given by a rational function:
\[
f_\mathrm{I}(p) = -
    \frac{\beta\,p^2-a\,p}{\alpha\,p^2 + \beta\,p + 1}.
\]
In this section, $\alpha$ and $\beta$ refer to mapping parameters,
not to the Twiss parameters.
In this simplified setting, the dynamics are governed by two
intrinsic quantities.
The first is the trace $a$, which determines the bare tune
$2\,\pi\,\nu_0 = \arccos(a/2)$.
The second is the ratio $\beta^2/\alpha$, which characterizes
the nonlinearity and directly influences the first twist
coefficient $\tau_0$, given by~\cite{zolkin2024MCdynamics,zolkin2024MCdynamicsIII}
\[
2\,\pi\,\tau_0 = -\frac{\alpha}{r_1\,r_2}\left[
    3\,r_4 + (a+8)\,\frac{r_3}{r_1}\,\frac{\beta^2}{\alpha}
\right].
\]

By introducing the small parameter $\epsilon$, we obtain the
following Taylor series of $f$:
\[
f(q) = a\,q - r_3\,\beta\,\epsilon\,q^2 +
\left(
r_3\,\beta^2 - r_4\,\alpha
\right)\,\epsilon^2\,q^3 + \ldots.
\]
From this expansion, we construct the approximate invariant:
\[
\begin{array}{l}
\K^{(n)}[p,q] = 
\cs +
\beta\,\Pi\,\Sigma\,\epsilon +
\left[ \alpha\,\Pi^2 + C_1\cs^2 \right]\epsilon^2    \\[0.3cm]
\,\,\,+\,2\,\beta\,C_1\Pi\,\Sigma\,\cs\,\epsilon^3   \\[0.3cm]
\,\,\,+\left[
    C_1\left(\beta^2\,r_2\,\Pi^3 + (2\,\alpha+\beta^2)\,\Pi^2\,\cs\right) +
C_2\,\cs^3 \right]\epsilon^4                         \\[0.3cm]
\,\,\,+\,\ob(\epsilon^5).
\end{array}
\]
While the first three terms match the exact invariant,
Eq.~(\ref{math:K-I}), all additional terms are proportional to
$C_i$.
The application of the averaging procedure ensures that for any
order $n>1$ and all $i > 0$, the coefficients $C_i$ must vanish
in order to satisfy $\dd I_n/\dd C_i = 0$.
Thus, this demonstrates that PT with averaging successfully
``passes'' its first test --- showing exact convergence once
the appropriate order is reached.

\subsection{Suris exponential mapping}

Next, we turn to the Suris mapping with the exponential invariant
(II).
As in the previous case, the invariant can be significantly
simplified by imposing a fixed point at the origin and requiring
$\K[0]=0$:
\[
\begin{array}{l}
\K_\mathrm{II}[p,q] =
        -a\,\sinh(p)\sinh(q)\\[0.3cm]
\qquad   +\,2\,\beta \,[\sinh(p+q)-\sinh(p)-\sinh(q)]     \\[0.3cm]
\qquad   +\,2\,\alpha\,[2\cosh(p)\cosh(q)-\cosh(p)-\cosh(q)].
\end{array}
\]
Furthermore, if $\alpha \neq 0$, the invariant can be rescaled so
that $\alpha = 1$, leaving only two intrinsic parameters.

\begin{figure*}[t!]
    \centering
    \vspace{1cm}
    \includegraphics[width=\linewidth]{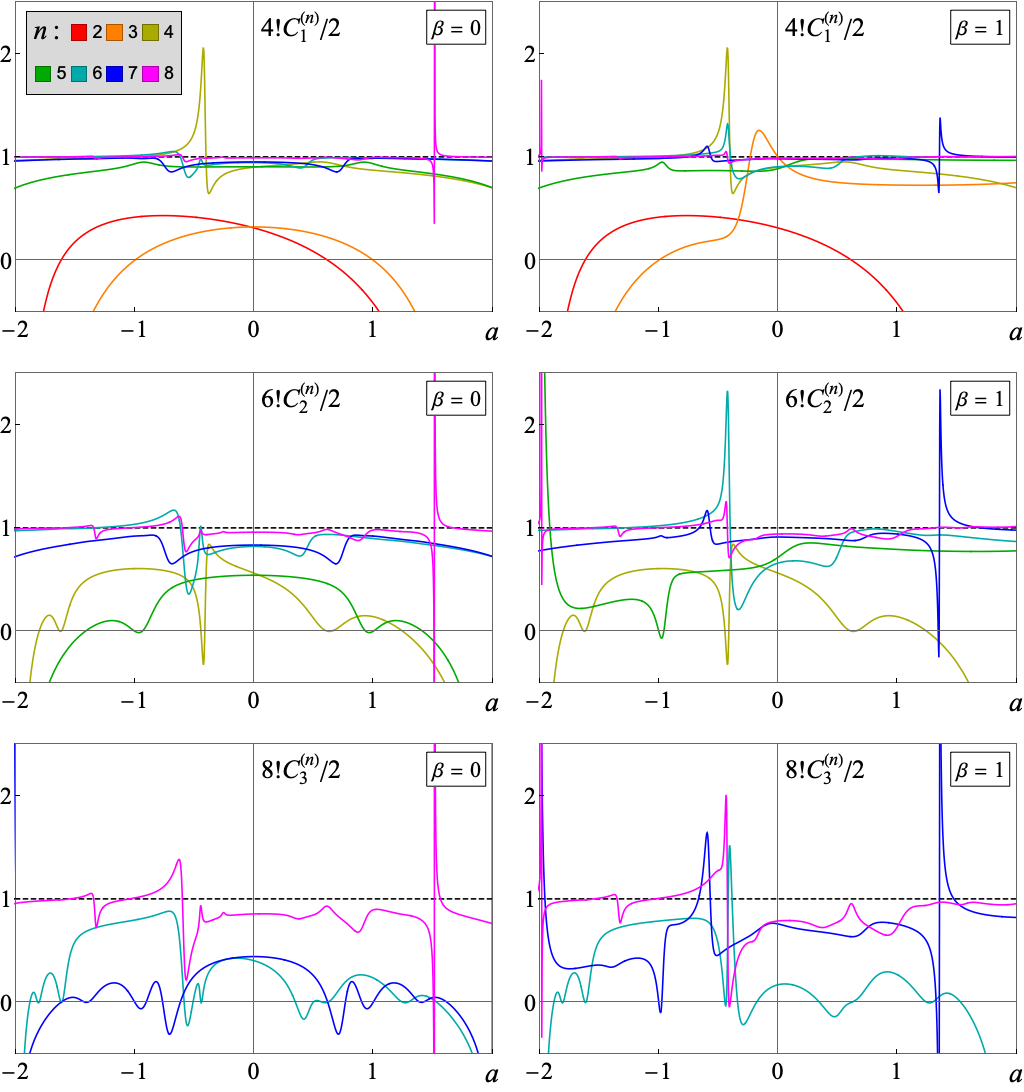}
    \caption{\label{fig:C123Suris}
    The first three coefficients $C_i^{(n)}$ ($i=1,2,3$) obtained
    using the averaging procedure for the Suris exponential
    mapping are shown as functions of the map parameter $a$
    (colored curves, as indicated in the legend).
    All coefficients are normalized as $(2(i+1))!C_i^{(n)}/2$
    and should be compared to the reference value of unity
    (represented by the dashed black line).
    The left and right columns display two sample cases:
    an odd mapping ($\beta = 0$) and a mapping with mixed
    nonlinearity ($\beta = 1$).
    As expected, for the odd mapping in odd orders, $C_i^{(n)}$
    are even functions of $a$.
    }\vspace{-1cm}
\end{figure*}

\newpage
$\,$
\newpage
$\,$

\newpage
In this form, the force function simplifies to
\[
f_\mathrm{II}(p)
= \log
\frac{(1+a/4)\,e^{2\,p}-(1-\beta)\,e^p+(1-\beta-a/4)}
     {(1+\beta-a/4)\,e^{2\,p}-(1+\beta)\,e^p+(1+a/4)},
\]
which we again expand in a Taylor series:
\[
a\,p - r_3\,\beta\,\epsilon\,p^2 +
    \left(
        r_3\,\beta^2 - r_4\,\frac{10-a^2}{12}
    \right)\epsilon^2\,p^3 + \ldots.
\]
Next, we compare the approximate invariant obtained before
applying the averaging procedure
\[
\begin{array}{l}
\ds \K_\mathrm{II}^{(n)}[p,q] = 
    \cs +
    \beta\,\Pi\,\Sigma\,\epsilon +
    C_1\cs^2\epsilon^2 +
    \frac{10-a^2}{12}\,\Pi^2\epsilon^2                   \\[0.35cm]
\ds \quad+\,
    \beta\,\frac{r_3\,\Pi+(24\,C_1-1)\,\cs}{12}\,\Pi\,\Sigma\,\epsilon^3 +
    C_2\cs^3\epsilon^4                                   \\[0.35cm]
\ds \quad+\,\Pi^2\epsilon^4\left[
    \beta^2\frac{(12\,C_1-1)(r_2\,\Pi+\cs)}{12}\right.   \\[0.45cm]
\ds \qquad\left.+\,
    \frac{(23-2\,a^2)(r_4\,\Pi-\cs)+60\,C_1(10-a^2)\,\cs}{6!/2}
\right]                                                  \\[0.55cm]
\ds \quad+\,\ob(\epsilon^5),
\end{array}
\]
with the power series expansion of the actual invariant
\[
\begin{array}{l}
\ds \K_\mathrm{II}[p,q] = 
    \cs +
    \beta\,\Pi\,\Sigma\,\epsilon +
    \frac{\cs^2}{12}\,\epsilon^2 +
    \frac{10-a^2}{12}\,\Pi^2\epsilon^2                   \\[0.35cm]
\ds \qquad+\,
    \beta\,\frac{r_3\Pi+\cs}{12}\,\Pi\,\Sigma\,\epsilon^3 +
    \frac{\cs^3}{360}\,\epsilon^4                        \\[0.35cm]
\ds \qquad+\,
\frac{r_4(23-2\,a^2)\,\Pi+3\,(9-a^2)\,\cs}{6!/2}\,\Pi^2\epsilon^4\\[0.5cm]
\ds \qquad+\,\ob(\epsilon^5).
\end{array}
\]
From this comparison, we extract the exact values of the
coefficients
\[
C_i = \frac{2}{[2(i+1)]!},
\qquad\qquad
i=0,1,2,\ldots.
\]
These coefficients are independent of the mapping parameters and,
unlike in the symmetric McMillan map case, do not vanish.
Moreover, summing all terms of the form $C_i\,\cs^{i+1}$ gives a
closed expression:
\[
    \sum_{i = 0}^\infty C_i\,\cs^{i+1} =
    2\,(\cosh\sqrt{\cs} - 1).
\]

Since coefficients $C_{i,j}$ can be determined exactly, our focus
remains on understanding $C_i^{(n)}$, which are obtained via the
averaging procedure.
In general, these coefficients depend on both $a$ and $\beta$.
For example,
\[
\begin{array}{l}
\ds \!C_1^{(2)} \!=\! \frac{5}{48}\,\frac{r_5}{r_1\,r_2},       \\[0.25cm]
\ds \!C_1^{(3)} \!=\!  \frac{r_3}{30\,r_2}
\frac{7\,r_1\,r_4^2\,r_6\,S_0\,S_1 + 6\,r_4\,S_2\,\beta^2 +
        180\,r_2\,r_3\,S_3\,\beta^4}
     {5\,r_1^2\,r_4^2\,S_0^2 - 24\,r_1\,r_3\,r_4\,S_0\,S_4\,\beta^2 +
        72\,r_3^2\,S_3\,\beta^4},
\end{array}
\]

\noindent
where

\begin{figure*}[t!]
    \centering
    \includegraphics[width=0.97\linewidth]{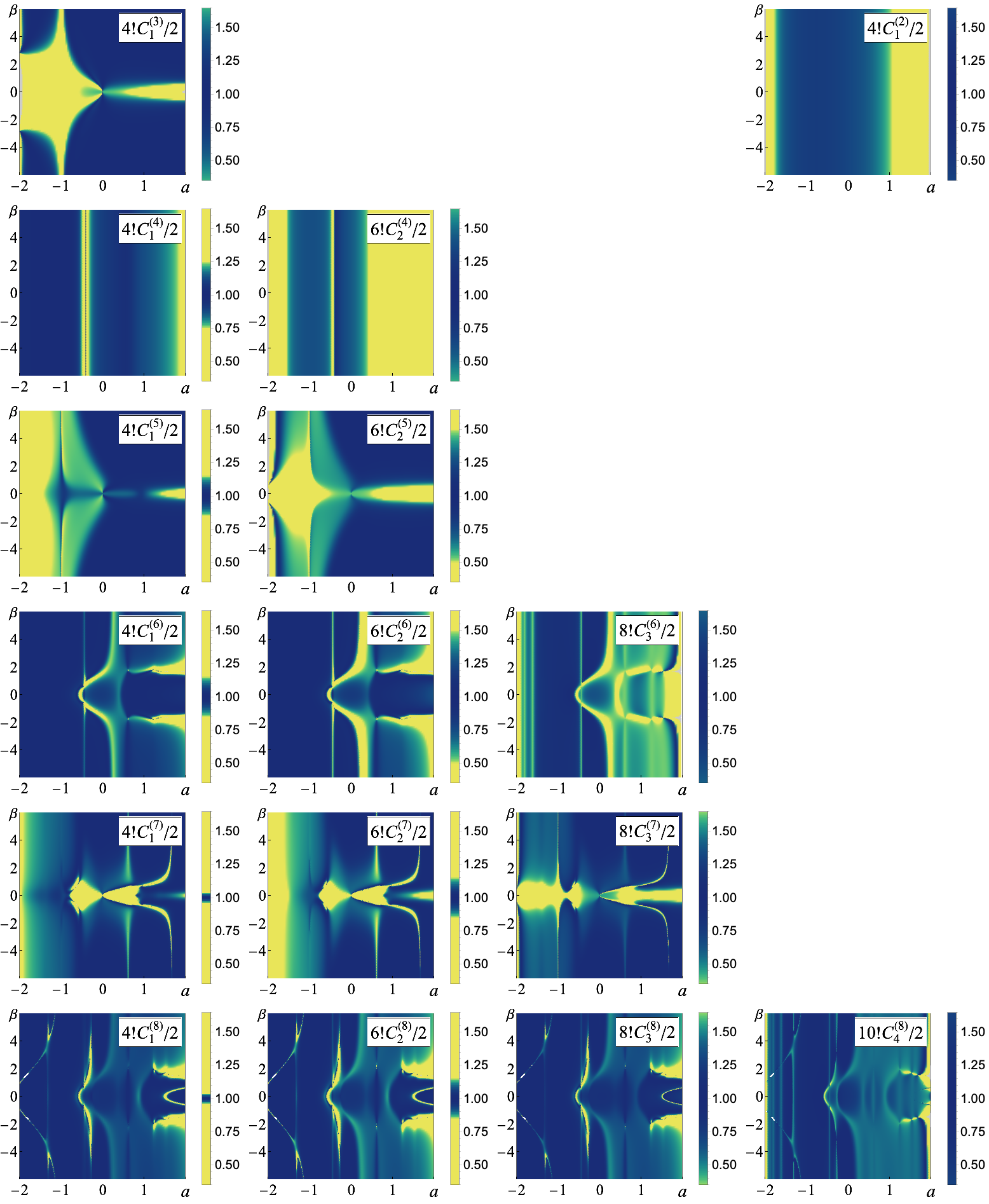}
    \caption{\label{fig:8ordSuris}
    Coefficients $C_i^{(n)}$ obtained using the averaging procedure
    for the Suris exponential mapping are shown as functions of the
    map's parameters, $a$ and $\beta$, with values represented by
    the color map.
    The different columns correspond to $i = 1, \ldots, 4$, while the
    rows represent $n = 3, \ldots, 8$;
    for a more compact presentation, the plot for $C_1^{(2)}$ is
    placed in the top right corner.
    All coefficients are normalized as $(2(i+1))!C_i^{(n)}/2$ and
    should be compared to unity (shown in dark blue).
    Note that for a given $i$, the color map is progressively
    rescaled with increasing $n$ to better illustrate convergence.
    }\vspace{-1cm}
\end{figure*}

\vspace{-0.5cm}
\[
\begin{array}{l}
\ds S_0 = 10 -    a^2,          \qquad\qquad\,\,\,\,
S_3 = 5 + 2\,a + 3\,a^2,        \\[0.2cm]
\ds S_1 = 23 - 2\,a^2,          \qquad\qquad
S_4 = 3 + 2\,a,                 \\[0.2cm]
\ds S_2 = 761 + 561\,a - 385\,a^2 - 315\,a^3 + 29\,a^4 + 24\,a^5.
\vspace{-0.2cm}
\end{array}
\]
Since $C_i^{(n)}$ are not required to obtain the power series 
$\nu(J)|_{J=0}$, which describes the dynamics in a simply connected
region around the origin, our goal is to attempt a reconstruction
of the full invariant of motion that extends globally.
To illustrate our findings, we refer to Figs.~\ref{fig:C123Suris}
and \ref{fig:8ordSuris}.

Fig.~\ref{fig:C123Suris} shows the coefficients $C_i^{(n)}$
obtained at different orders ($n \leq 8$, shown as colored curves)
as functions of the trace parameter $a$.
A uniform scaling is applied to all curves so that the value of
$C_i$ is normalized to unity (dashed black line).
Each row corresponds to a different $i$, displaying $C_1^{(n)}$,
$C_2^{(n)}$ and $C_3^{(n)}$, respectively.
The two columns represent different values of the nonlinear
parameter: the left column, $\beta = 0$, represents an odd-force
function, while the right column, $\beta = 1$, reflects a mixed
(odd and even) nonlinearity.
The plots show that although $C_i^{(n)}$ depends on both
parameters, it increasingly approximates a constant value
as $n$ grows, approaching unity in the expected limit.

One key difference from the previously considered
integrable McMillan map is that, in this case, convergence is
not expected at any finite order.
Instead, the behavior suggests that the infinite series sum
appears to {\it converge almost everywhere}, resembling the
Gibbs phenomenon observed in Fourier series for square,
triangular, and other discontinuous functions.

\vspace{0.1cm}\noindent
{\bf Observation}.
Notably, none of the coefficients exhibit singularities within
the parameter range corresponding to a stable fixed point at the
origin ($|a|<2$).
This aligns with expectations, as Suris mappings lack nonlinear
resonances, except for special degeneracies where $r_3 = 0$ or
$r_4 = 0$;
this stands in contrast to generic chaotic systems, where almost
each rational value of $\nu_0$ corresponds to a resonance at the
origin --- either a singular one, such as a touch-and-go
bifurcation, or a non-singular, e.g., $k$-island chain.
Moreover, as the order increases, errors become increasingly
localized, and their width progressively narrows.

Fig.~\ref{fig:8ordSuris} further explores this behavior by
mapping $C_i^{(n)}$ across both parameters $a$ and $\beta$.
Rows correspond to different orders $n=2,\ldots,8$,
while columns represent different values of $i=1,\ldots,4$.
The same rescaling as in Fig.~\ref{fig:8ordSuris} is applied,
with dark blue indicating a reference value of 1.
The convergence can be observed as, (i) for a given $i$ and
increasing $n$, the regions of large error, highlighted in
yellow, become increasingly localized, while simultaneously,
(ii) the color map undergoes rescaling, reflecting an
improved overall approach to unity.
With that, we consider another test to be ``passed''
--- albeit in the limit $n \to \infty$.
The results for the trigonometric Suris mapping exhibit
qualitatively similar behavior, as the two cases are inherently
related.

\newpage
$\,$
\newpage
$\,$

\newpage
\section{\label{sec:Chaos}Applications to chaotic systems}

In this section, we explore various applications of our
perturbative approach using the chaotic quadratic and cubic
H\'enon maps  \cite{henon1969numerical}.
Owing to the wealth of existing analytical results (see, e.g.,
\cite{dulin2000henon,dullin2000twistless,sterling1999homoclinic}),
these systems serve as an excellent test ground.
Both transformations can be written in McMillan form, with force
functions given by $f(p) = a\,p + p^2$ and $f(p) = a\,p + p^3$
respectively.

We begin with a few specifically selected examples and discuss the
approximation of the nonlinear rotation number.
We then extend our study to the joint space of phase-space variable
and mapping parameter, using our perturbative technique to describe
the intricate structure of the fractal stability diagram.
Afterward, we subject the square matrix (SM) method to the same test,
enabling a direct comparison.

Finally, in Appendix~\ref{secAPP:PT}, we present the level sets
of the approximate invariants for all three cases discussed in
Subsection~\ref{sec:NuQ}.
These are intended for readers' independent exploration, although
we include brief commentary and a complementary set of figures
produced using the SM method, Appendix~\ref{secAPP:SM}.

\vspace{-0.2cm}
\subsection{\label{sec:NuQ}Nonlinear rotation number}

\vspace{-0.1cm}
As the rotation number is an intrinsic dynamical quantity, its
accurate evaluation remains a central challenge.
In the nearly (quasi-)integrable example discussed at the end of
the first part of this series (Subsection II E, Fig. 14), we
showed how $\nu(J)$ can be extracted under favorable conditions.
However, in more realistic scenarios --- where the phase space
features prominent island chains and extensive chaotic regions ---
the action variable becomes less practical to use.
Moreover, evaluating the action numerically is non-trivial,
complicating meaningful comparisons between analytical and
numerical approaches.

A partial remedy comes from considering the rotation number as
a function of the phase space coordinates themselves, rather
than $J$.
In this formulation, we can directly compute $\nu(\K[p,q])$ from
the approximate invariants via Danilov's theorem.
Furthermore, numerical orbit tracking naturally yields
$\nu[p_0,q_0]$ as a function of initial conditions, facilitating
direct comparison.
In accelerator physics, this approach aligns well with practical
measurements, where the nonlinear betatron tune is inherently
observed as a function of the particle's position and momentum.

\begin{figure*}[t!]
    \centering
    \includegraphics[width=\linewidth]{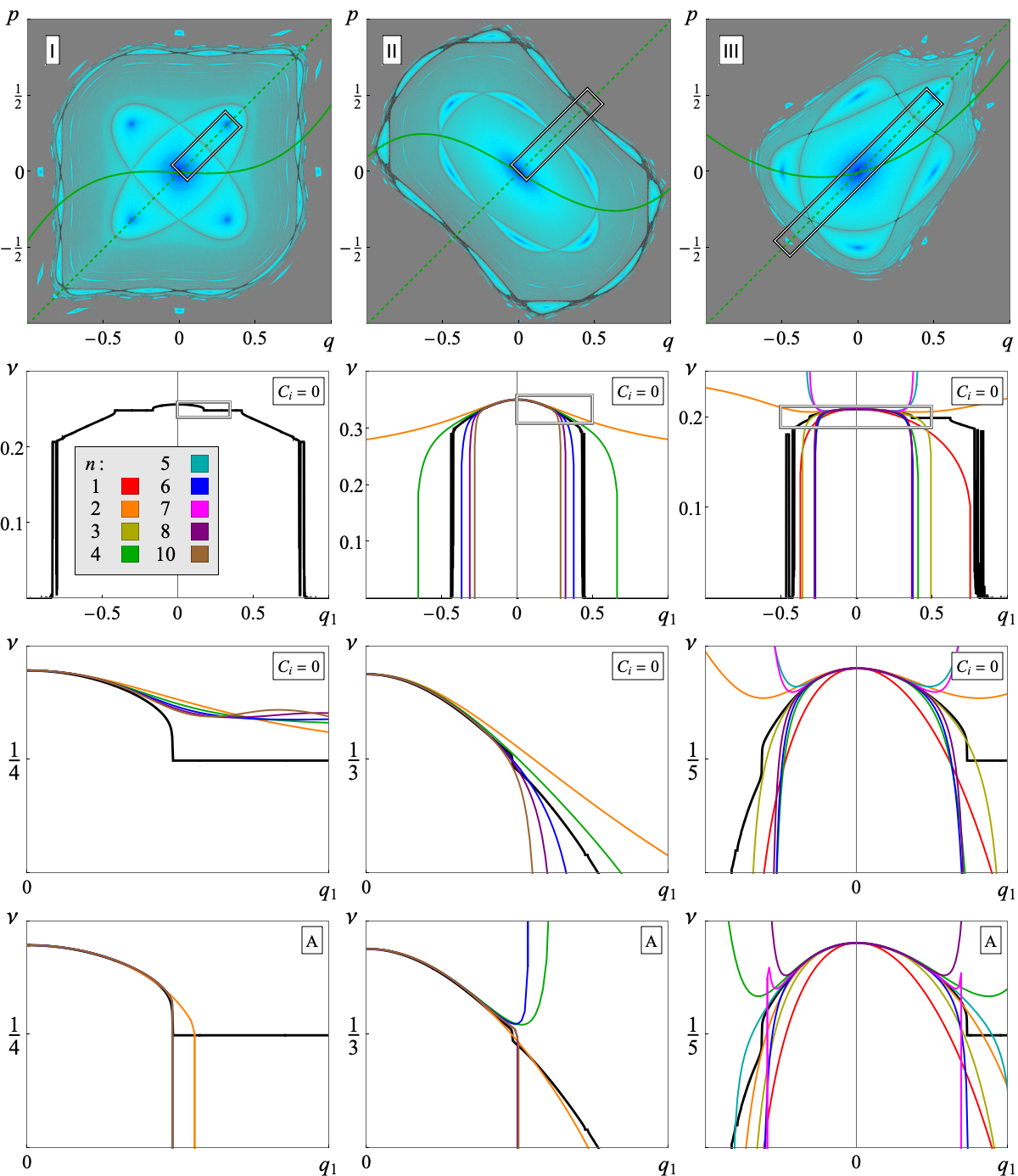}
    \caption{\label{fig:NuQExamples}
    Top row shows phase space portraits highlighted with the REM
    indicator for the H\'enon cubic (I, II) and quadratic (III)
    mappings, corresponding to trace parameter values $a =-1/10$,
    $-6/5$, and $1/2$, respectively.
    Dashed and solid green curves indicate the first and second
    symmetry lines.
    Second row presents rotation number $\nu(q_1)$ as a function
    of the initial coordinate along the first symmetry line,
    $\z_0=(q_1,q_1)$,
    obtained from numerical orbit tracking (black thick curves)
    and from non-averaged approximate invariants with $C_i = 0$
    (colored curves, labeled in the legend).
    Both rows include a white rectangle marking the region selected
    for magnification.
    Two bottom two rows contains close-up views of the selected
    region, showing the rotation number curves extracted from
    non-averaged invariants (third row) and after applying the
    averaging procedure (A) (fourth row).
    }\vspace{-1cm}
\end{figure*}

Fig.~\ref{fig:NuQExamples} presents three examples drawn from
the family of He\'non cubic (I, II) and quadratic (III) maps.
Here, we consider non-resonant bare tune values $\nu_0$, setting
these cases apart from those in Subsection II D of the first part.
The values of the trace parameter $a$ are chosen to lie near
low-order resonances --- $1/4$, $1/3$, and $1/5$ for cases I
through III, respectively --- so that the island chains can still
be captured by relatively low-order perturbative expansions.

The top row of Fig.~\ref{fig:NuQExamples} shows phase space
portraits colored using the REM indicator, which highlights
resonance structures: dark blue corresponds to quasi-linear
motion, cyan to quasi-integrable, and gray to chaotic or
unstable initial conditions.
Below, we plot the rotation number as a function of the initial
coordinate along the first symmetry line $l_1$, i.e.,
$(q_0,p_0) = q_1(1,1)$, using both numerical orbit tracking
(black thick curves) and results extracted from non-averaged
approximate invariants (with $C_i = 0$) and averaged (denoted A).
The second row presents the rotation number on the same scale as
the corresponding phase space plots, while the two bottom rows
show zoomed-in regions (highlighted with white rectangles) for
better clarity.

Focusing on these magnified views, we again observe the clear
advantage of the averaging procedure.
In case I, for example, non-averaged approximations fail to
capture the separatrix associated with the $1/4$ island chain
(which appears as a plateau at $\nu = 1/4$), despite correctly
predicting the twist coefficient.
In contrast, the averaged second-order curve (orange) already
provides a reasonable approximation, while higher-order terms
further improve the result --- accurately reproducing the steep
frequency drop near the separatrix, which often resembles a
local collapse.
This is precisely the situation where low-order approximations
$\nu_k$ offer a better qualitative and quantitative description
than both the non-averaged case and the Lie power series $\nu(J)$.

In cases II, III the benefit of averaging remains clear.
A comparison of the second-order curves (orange) in both cases
shows that averaging produces significantly more accurate results.
In particular, for case II, the curves corresponding to orders
8 (purple) and 10 (brown) closely match the numerically computed
rotation number inside the entire 6-island chain (comprising a
pair of 3-cycles).

As a further check, we include in Appendix~\ref{secAPP:PT} the
level sets of the averaged invariants used to compute the rotation
numbers shown here --- see Figs.~\ref{fig:CaseI}, \ref{fig:CaseII},
and \ref{fig:CaseIII} for completeness.
In case I, we observe that for $n \geq 4$, the PT provides a
highly accurate description of the inner chain of four islands.
As the order $n$ increases, additional features corresponding to
higher-period orbits begin to emerge.
Although these features may appear to ``fluctuate'' between
successive orders, the overall boundary of stability is captured
reasonably well.
In case II, the situation is similar: the chain of six islands
(representing two groups of 3-cycles) becomes clearly resolved
at $n=8$.
While PT again begins to break down at larger amplitudes, as in
case I, the relatively low-order approximation not only captures
the island chain but also succeeds in extending beyond the
separatrix by a non-negligible distance.
It is worth noting that, while we focus here on the simply
connected region inside the largest inner chain, Danilov's
theorem can be applied to the outer level sets surrounding
the origin as well.
Case III presents the most challenging scenario.
Even so, the $n=7$ approximation yields interesting results,
successfully revealing all of the islands.

\newpage
$\,$
\newpage
$\,$

\begin{figure*}[t!]
    \centering
    \includegraphics[width=\linewidth]{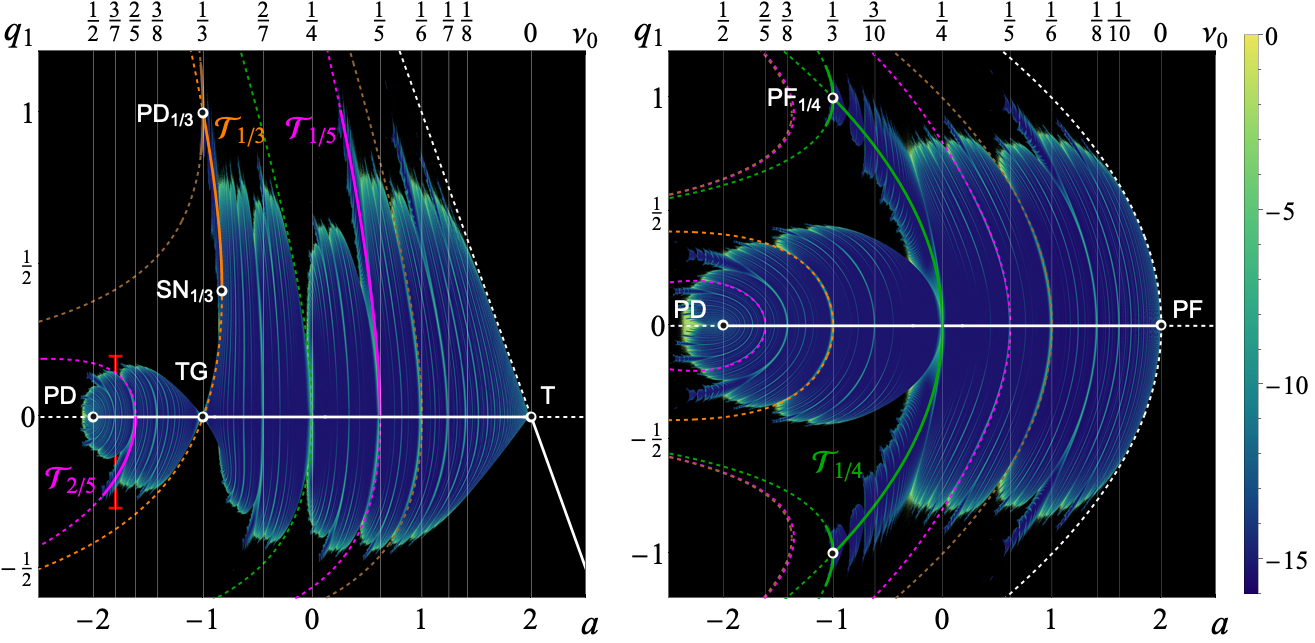}\vspace{-0.4cm}
    \caption{\label{fig:FMAall}
    Isochronous diagrams for the H\'enon quadratic $f(p)=a\,p+p^2$
    (left) and cubic $f(p)=a\,p+p^3$ (right) mappings.
    Each diagram displays the FMA index for stable initial
    conditions at different values of the trace parameter
    $a$ across the coordinate $q_1$, measured along the first
    symmetry line $l_1$.
    The top axis provides corresponding values of the bare
    rotation number $\nu_0$ for select low-order resonances.
    Additional curves indicate the locations of isolated fixed
    points (white) and periodic $n$-cycles ($n \leq 6$), with
    solid for stable and dashed for unstable.
    Specifically, orange represents 3-cycles, green denotes 4-cycles,
    magenta corresponds to 5-cycles, and brown marks 6-cycles.
    Key bifurcations associated with these points are labeled with
    Latin letters: transcritical (T), pitchfork (PF), period doubling
    (PD), saddle-node (SN), and touch-and-go (TG).
    Arnold tongues, associated with some $k$-island chain bifurcations,
    are marked as $\mathcal{T}_{\nu_0}$.
    The red line segment at $\nu_0 = 3/7$ (quadratic map) serves
    as a reference for Fig.~\ref{fig:FMAPhSp}.
    }\vspace{-0.5cm}
\end{figure*}

\newpage
\subsection{\label{sec:Simply}
Simply connected region and chains of islands}

\begin{figure}[h!]
    \centering
    \includegraphics[width=0.922\linewidth]{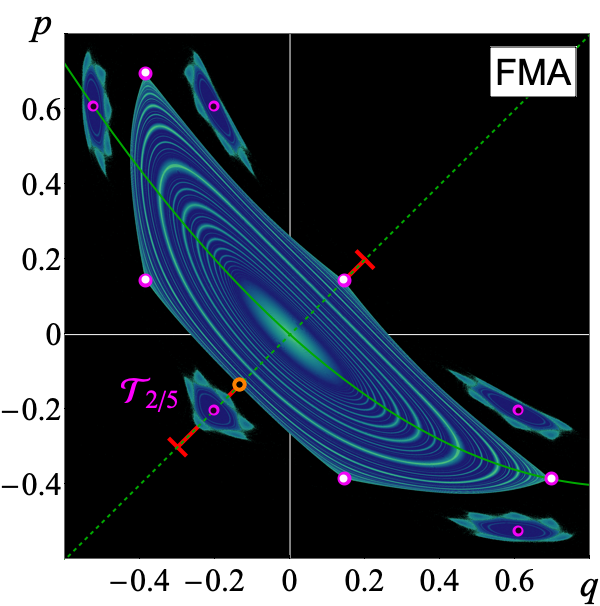}\vspace{-0.3cm}
    \caption{\label{fig:FMAPhSp}
    Phase space diagram for the H\'enon quadratic map at
    $\nu_0 = 3/7$.
    Magenta points indicate 5-cycles, with black-dot markers
    representing stable centers associated with a 5-island chain
    ($\mathcal{T}_{2/5}$) and hollow markers indicating unstable
    solution.
    Green curves represent the first (dashed) and second (solid)
    symmetry lines.
    The red line segment corresponds to Fig.~\ref{fig:FMAall}.
    }
\end{figure}

\vspace{-0.2cm}
One challenge in analyzing phase space plots lies in the choice of
parameters.
A ``carefully'' selected case study can create the illusion that a
perturbation method is remarkably effective, while, conversely,
cases where convergence can not be achieved (or has not yet been
approached) may give the false impression that the method is
unreliable.
To mitigate this issue, we extend our study of H\'enon mappings
beyond individual phase space plots, shifting our focus to a mixed
space representation that combines coordinates along a symmetry
line with variations in the mapping parameter $a$.
This broader perspective helps reveal underlying structures that
may not be apparent in isolated phase space views.
We refer to these diagrams as {\it isochronous} (along $l_1$) and
{\it period-doubling} (along $l_2$), as they cut through all fixed
points and 2-cycles, respectively.
A dedicated manuscript~\cite{zolkinHenonSet} explores this topic
in greater detail.
Fig.~\ref{fig:FMAall} presents isochronous diagrams for the H\'enon
quadratic and cubic mappings, while Fig.~\ref{fig:FMAPhSp} provides
a corresponding phase space diagram, with the red line segment
serving as a visual cue for comparison.
The color scale represents the natural logarithm of the FMA index,
capturing numerical diffusion in the rotation number:
$\log_{10}(10^{-16} + |\nu_I-\nu_{II}|)$
where $\nu_{I,II}$ are the rotation numbers computed over the first
and second halves of the iteration window.

\newpage

\begin{figure*}[t!]
    \centering
    \includegraphics[width=\linewidth]{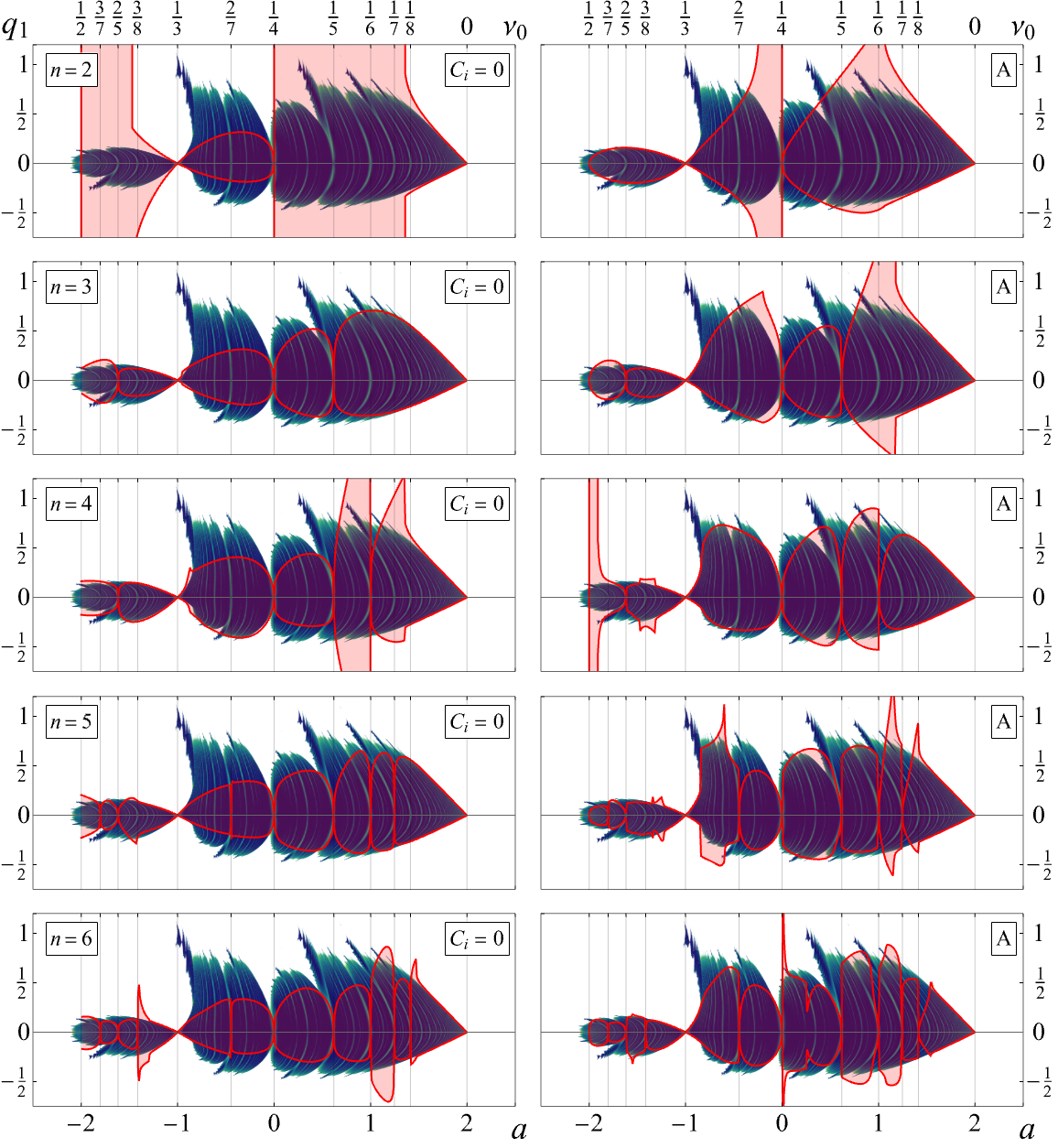}\vspace{-0.3cm}
    \caption{\label{fig:BoundariesSX}
    Simply connected region (red) along the first symmetry line, 
    extracted from non-averaged approximate invariants with
    $C_i=0$ and averaged invariants $\langle\K^{(n)}\rangle$ (A),
    overlaid on the isochronous diagram for the H\'enon quadratic
    map, $f(p) = a\,p + p^2$.
    }\vspace{-0.28cm}
\end{figure*}

\newpage

\vspace{-0.1cm}
In this subsection, we explore the potential use of our
approximate invariants to identify the locations of island
chains.
This question is particularly relevant in accelerator physics,
where a detailed understanding of phase space is essential
for improving the dynamic aperture --- the region of stable
trajectories around the origin or reference orbit.
A larger dynamic aperture enables better particle capture and
reduced beam losses, ultimately enhancing machine performance.
Moreover, this knowledge aids in optimizing lattice designs for
resonant extraction~\cite{zolkin2024MCdynamics}.

\begin{figure*}[t!]
    \centering
    \includegraphics[width=\linewidth]{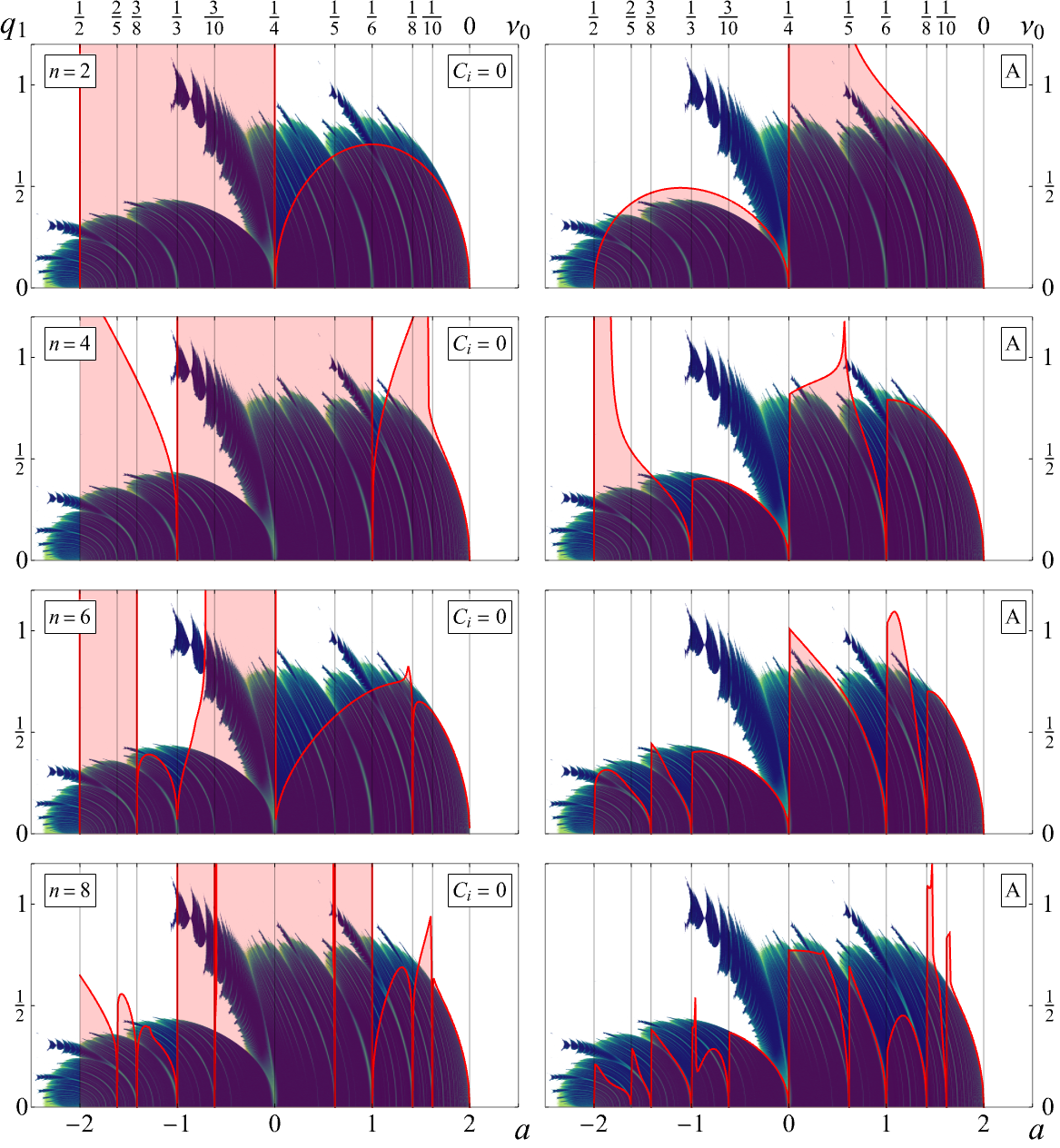}\vspace{-0.4cm}
    \caption{\label{fig:BoundariesOC}
    Same as Fig.~\ref{fig:BoundariesSX}, but for the H\'enon cubic
    map, $f(p) = a\,p + p^3$.
    }\vspace{-0.6cm}
\end{figure*}

\newpage
To investigate this, we analyze both the averaged approximate
invariant $\langle\K^{(n)}\rangle[p,q]$ and the non-averaged
version $\K^{(n)}[p,q]$, where we once again set all $C_i$ to
zero to maintain consistency with prior calculations.
From each, we extract the size of the simply connected region
around the origin along the first symmetry line, $l_1$.
This region is bounded by a {\it separatrix} --- an invariant
level set passing through a critical point, corresponding to an
unstable fixed point (cycle) with associated invariant manifolds.
Depending on the geometry, the separatrix may intersect a
symmetry line either directly or through a saddle point
(see \cite{zolkinHenonSet} for details).
To visualize some of these concepts, we refer to
Fig.~\ref{fig:FMAPhSp}.
In this phase space plot, the dynamic aperture is roughly
delineated by the unstable 5-cycle.
The associated separatrix is pierced by the symmetry line $l_1$
(green dashed): 
on the right through one of the cycle points, and on the left
directly through the unstable manifolds (marked by the orange
point).

Within this region, the dynamics appear quasi-integrable in the
sense that no prominent island chains or large chaotic zones are
visible at the given scale.
However, unlike the previously discussed REM-highlighted phase
space in Fig.~\ref{fig:NuQExamples}, this quasi-integrable region
exhibits numerous bands of lighter color.
While these may not resemble island chains at first glance, they
are indeed associated with orbits of rational rotation numbers,
and the corresponding mode-locked regions (Arnold tongues) become
visible upon closer magnification.
While REM (along with SALI/GALI or Lyapunov exponent-based methods)
is better suited for detecting chaos and other phase space features
such as twistless orbits~\cite{zolkinHenonSet}, the FMA indicator
is more sensitive to resonant structures such as hyperbolic points,
which justifies its use in this analysis.

In chaotic systems, a truly simply connected region does not
exist, since island chains emerge at every scale.
However, its boundary is well-defined within our approximate
invariants.
Figs.~\ref{fig:BoundariesSX} and \ref{fig:BoundariesOC} provide
the extracted boundary (shown as a red envelope curve)
superimposed onto the isochronous diagrams from
Fig.~\ref{fig:FMAall}.
If the fixed point at the origin is the only critical point of
the invariant, no bounding separatrix exists and the dynamic
aperture is effectively infinite;
in such cases, no red curve is shown, and stability is indicated
by the semi-transparent fill extending vertically.

Comparing the left-hand plots (non-averaged invariants with
$C_i = 0$) to the right-hand plots (averaged invariants), we
observe an increasing number of ``pinchings'' in the red envelope
as  $n$ increases.
These collapses of the simply connected region correspond to
resonances that appear in the expected orders.
For general maps, including the quadratic H\'enon map, the set of
resonance values $\{\nu_r^{(n)}\}$ appearing at order $n$ are
given by the left half of the Farey sequence:
\[
\{\nu_r^{(n)}\} = \left\{\frac{l}{m}  \in F_{n+2} \,\, \bigg|\,\,
                  0 \leq \frac{l}{m} \leq \frac{1}{2}\right\}.
\]
In the case of odd mappings, such as H\'enon cubic map, the even
resonances $r_{2\,k}$ first appear at their corresponding even
orders $n=2\,k-2$, while odd resonances $r_{2\,k+1}$ arise along
with even resonances of doubled period $r_{4\,k+2}$ at order
$n=4\,k$:
\[
\begin{array}{l|ll}
n &\,\,f(p)                            &\quad f_\mathrm{odd}(p)             \\\hline
0 &\,\,r_1,r_2                         &\quad (r_2,r_1)                     \\
1 &\,\,r_1,r_2,r_3                     &\quad                               \\
2 &\,\,r_1,r_2,r_3,r_4                 &\quad (r_2,r_1),r_4                 \\
3 &\,\,r_1,r_2,r_3,r_4,r_5             &\quad                               \\
4 &\,\,r_1,r_2,r_3,r_4,r_5,r_6         &\quad (r_2,r_1),r_4,(r_6,r_3)       \\
5 &\,\,r_1,r_2,r_3,r_4,r_5,r_6,r_7     &\quad                               \\
6 &\,\,r_1,r_2,r_3,r_4,r_5,r_6,r_7,r_8 &\quad (r_2,r_1),r_4,(r_6,r_3),r_8
\end{array}
\]

A more detailed comparison shows that, while both approaches (with
and without averaging) correctly identify the resonance locations,
averaging once again yields significantly more accurate results.

Each plot in Figs.~\ref{fig:BoundariesSX} and \ref{fig:BoundariesOC}
is rich in information, and we encourage the reader to explore
them in detail.
To observe convergence, one can select a low-order resonance
(using the complimentary top axis for $\nu_0$) and locate the
corresponding structure in the diagram --- e.g., cyan lines
corresponding to direct crossing of the separatrix, or Arnold
tongues indicating island chains.
By following the red envelope across increasing orders $n$, in
case of averaging procedure, one sees it more closely conform to
the resonance structure until it eventually ``switches'' to a
higher-order resonance.
Through systematic examination of these features, the advantage
of the averaging procedure becomes clear.

\subsection{\label{sec:SM}
Square Matrix method}

In this closing subsection, we compare our perturbative technique
with the Square Matrix (SM) method [{\bf REF?}].
The SM approach begins by reducing the dimensionality of the
coefficient matrix.
Then, through a Jordan decomposition, it derives a new set of
approximate action-angle variables and an associated invariant.
The deviation of this approximate action from constancy serves
as a measure of stability and frequency fluctuation --- analogous
to the approximate invariance in our method.

To structure the comparison, we revisit the criteria that guided
the development of our perturbative approach.
In the first part of this manuscript, we focused on local
convergence near the fixed point.
Two central requirements were:
(i) the rotation number should agree with its Lie series
expansion at $J = 0$, and
(ii) the method must remain valid even when the rotation number
is rational --- that is, at resonances.
As Lie algebra methods are known to match numerical results near
the origin, any approach failing to satisfy condition (i) should
be considered inadequate.
Throughout this work, we have demonstrated that there are
infinitely many ways to construct approximate invariants
(for $n \geq 2$) that fulfill this requirement to a given
order.
The SM method appears to meet this criterion, which makes
it a natural and interesting candidate for comparison.
However, when it comes to condition (ii), our understanding is
that the SM method is not inherently resonant.
We were unable to derive approximate invariants using SM that
resemble the resonant constructions discussed in the first part
(see Subsection II D and Appendix~4, which includes the resonant
normal forms).
This suggests that the SM method lacks built-in mechanisms for
treating resonant dynamics analytically.

\begin{figure*}[t!]
    \centering
    \includegraphics[width=\linewidth]{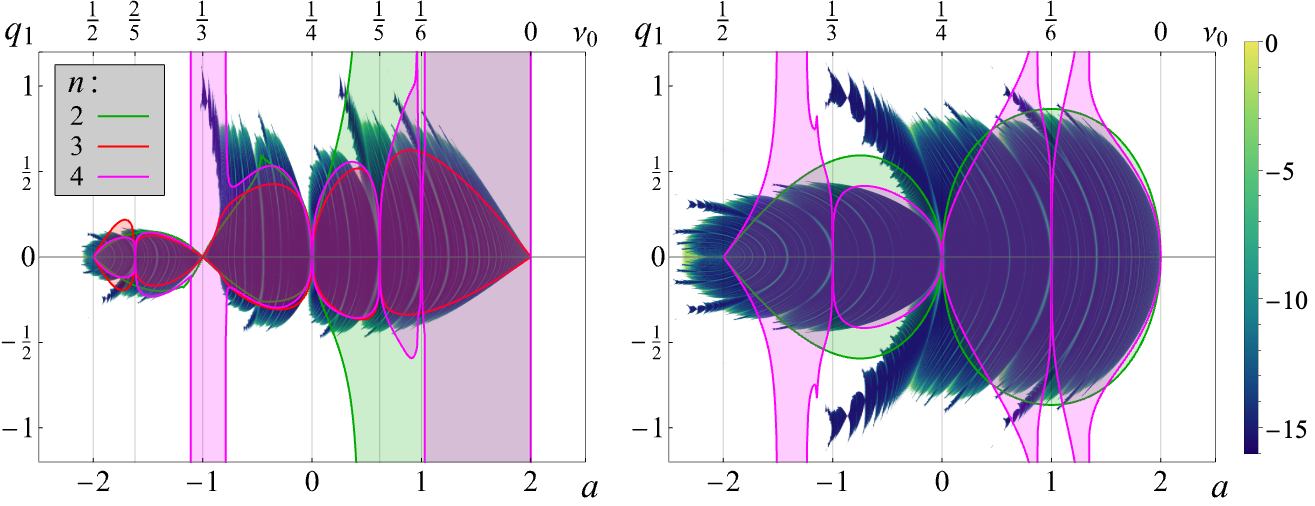}
    \caption{\label{fig:BoundariesSM}
    Same as Figs.~\ref{fig:BoundariesSX} (left plot) and
    \ref{fig:BoundariesOC} (right plot), but showing the boundaries
    of the simply connected region derived from the square matrix
    approximate invariants.
    Results from all available orders are displayed together, with
    each order indicated by a distinct color, as shown in the legend.
    Here, $n$ denotes the perturbative order matching our PT method,
    while the matrix size is given by $n + 2$.
    }\vspace{-0.15cm}
\end{figure*}

The second set of criteria, relevant to large-amplitude dynamics
and explored in the present part, includes:
(iii) the ability to recover the exact invariant for integrable
systems, and
(iv) the capacity to capture (or even surpass) the island chains
characteristic of chaotic systems.
Our attempt to analytically recover the invariant for the
integrable McMillan map using the SM method failed in the
same way our own approach fails when the averaging procedure
is not applied: the invariant, instead of terminating properly
at second order (as it should), grows with $n$ and displays
spurious resonant denominators.
Nevertheless, we still proceed to apply the SM method to chaotic
systems for comparison.
Despite its limitations in the integrable regime, it remains
worthwhile to assess how well SM captures features like island
chains in more complex dynamical settings.

Fig.~\ref{fig:BoundariesSM} revisits the isochronous diagrams
for the quadratic and cubic H\'enon maps, this time overlaying
them with the simply connected regions extracted from the SM
approximations.
These are shown for orders corresponding to $n = 2,\,3,\,4$ in
our perturbative (PT) method.
We omit the $n=1$ result to avoid redundancy, as it coincides
exactly with the PT method without averaging and with vanishing
$C_1$ coefficient.

While our approximate invariants were obtained analytically for
all orders, in the SM method we were only able to derive explicit
analytical expressions for the lower orders shown.
Attempts to extend the SM method numerically for this particular
test were unfortunately unsuccessful.
In our experience, as one approaches resonant values of the bare
tune $\nu_0$, the numerical procedure becomes unstable, with
resonant denominators diverging or ``exploding,'' leading to
contamination of the plots with numerical artifacts.
This limitation may reflect challenges in the our implementation
rather than the SM method itself.
For this reason, we present only the most reliable results.

Overall, the qualitative behavior of the SM method closely
parallels our PT results without averaging: as the order $n$
increases, both methods consistently reveal resonances,
following a Farey-like sequence.
This is seen in the form of singularities or zero-crossings
in the boundary of the simply connected region as $\nu_0$
approaches rational values.
However, as the tune shifts away from resonance, the curves
representing the SM-inferred regions increasingly deviate from
the expected light-colored structures revealed by the FMA.
Closer inspection of the diagrams confirms that PT with averaging
consistently provides better global estimates of the boundary
structure among the three methods.
That said, even the averaged method is eventually limited by the
persistent switching of dominant island chains, which ultimately
reduces the size of the simply connected region.

Figs.~\ref{fig:CaseI-IIsm} and \ref{fig:CaseIII} in Appendix
\ref{secAPP:SM} explore the SM method further by applying it to
the specific case studies I -- III introduced earlier in this
section;
in these examples, we were able to obtain numerically stable
results up to $n = 10$.
In case I, the SM method reveals the inner portion of the
separatrix associated with the 4-island chain.
Beginning at $n=6$, the approximation becomes highly accurate ---
comparable to the PT result at $n=4$ with averaging, which already
fully captures all four islands.
For case II, the SM method again starts to approximate the inner
part of the island chain from $n=6$ onward.
However, even at $n=10$, the method still noticeably deviates from
the true locations of the unstable 3-cycles.
In contrast, PT method with averaging reveals all six islands by
$n=6$, and at $n \geq 14$ the approximation becomes nearly
indistinguishable from the exact structure.
In the more challenging case III, the pattern is similar: from
$n \geq 3$ the SM method begins to capture the inner region of
the 5-island chain, gradually converging toward the correct
geometry.
Yet here too, the PT method with averaging appears to reach the
true locations of the chain nodes more rapidly and with greater
consistency;
however, it is important to note that even our method encounters
difficulties when it comes to consistently resolving fine island
structures.

\newpage
In all three cases, as $n$ increases, the SM method seems to
approach the dynamics of an integrable twist map inside the
separatrix --- but struggles to go beyond it and resolve structures
outside that region.
While a comprehensive study of the convergence properties of each
method is beyond the scope of this work, our current observations
suggest that PT with averaging tends to deliver more accurate
predictions.

\vspace{0.3cm}
\section{\label{sec:Conclusion}Conclusions and possible generalizations}

In this second part of the manuscript, we expanded our perturbative framework beyond the local dynamics near the fixed point to evaluate
its performance across a broad range of nonlinear systems, including
chaotic maps.
Our primary focus was on the global structure of phase space and the
ability of the method to describe key features such as rotation number,
island chains, and approximate invariants with high fidelity.

The analysis was guided by two additional criteria, building upon
the foundational requirements established in the first part.
Specifically, we evaluated:
(iii) whether the method can recover exact invariants in integrable
systems, and
(iv) whether it can accurately capture --- and in some cases surpass
--- the island chains that dominate the phase space of chaotic
systems.

Both objectives were met.
For integrable systems, we demonstrated that the perturbative
expansion, when properly averaged, (almost everywhere) converges
to the exact invariants.
In the non-integrable regime, we showed that the method
successfully resolves nonlinear resonance structures, including
island chains, and tracks their evolution across parameter space.
The technique consistently produced reliable boundaries of the
simply connected stable region.
Notably, the rotation number --- computed from the approximate
invariant via Danilov's theorem --- closely follows the numerical
tune extracted from orbit tracking across a wide range of initial
conditions.

To place our method in context, we also compared it to the Square
Matrix (SM) approach.
While SM shares some formal similarities --- particularly at low
order --- and seems to satisfy the essential
requirement of local
convergence near the fixed point, we found that it lacks a built-in
mechanism for resonant expansions and struggles to recover the
correct invariant in integrable systems without divergence.
In chaotic systems, SM does resolve inner parts of island chains
with increasing order, but it tends to underestimate their extent
and fails to surpass separatrices.
In contrast, our perturbative method with averaging consistently
provided sharper boundaries and better alignment with true
dynamical structures, especially near resonances.

The efficiency and generality of the method are further underscored
by its application to several realistic accelerator lattices in the
final part of this series.
All accelerator-related plots in this work were derived from a single
unified object --- the approximate invariant --- which depends
parametrically on the map coefficients.
This demonstrates the method's versatility and practicality as a
diagnostic and design tool in beam dynamics.
Despite the complexity of the underlying maps, the analytic
construction of the invariant required only a modest number
of perturbative terms, while averaging systematically improved
both accuracy and convergence.

In summary, we have developed a perturbative technique that not
only satisfies key theoretical benchmarks, but also performs
remarkably well across integrable and chaotic regimes.
It offers a transparent, analytic, and efficient framework for
exploring nonlinear dynamics in symplectic maps, and provides
a valuable complement to existing methods in accelerator physics
and beyond.


\section{Acknowledgments}

The authors would like to thank Taylor Nchako (Northwestern
University) for carefully reading this manuscript and for her
helpful comments.
S.N. work is supported by the U.S. Department of Energy,
Office of Science,
Office of Nuclear Physics under contract DE-AC05-06OR23177.
I.M. acknowledges that his work was partially supported by the
Ministry of Science and Higher Education of the Russian Federation
(project FWUR-2025-0004).
S.K. is grateful to his supervisor, Prof. Young-Kee Kim
(University of Chicago), for her valuable mentorship and continuous
support.

\appendix

\newpage
\onecolumngrid
\section{\label{secAPP:PT}Averaged approximate invariants for
case studies I -- III}

\begin{figure*}[h!]
    \centering
    \includegraphics[width=\linewidth]{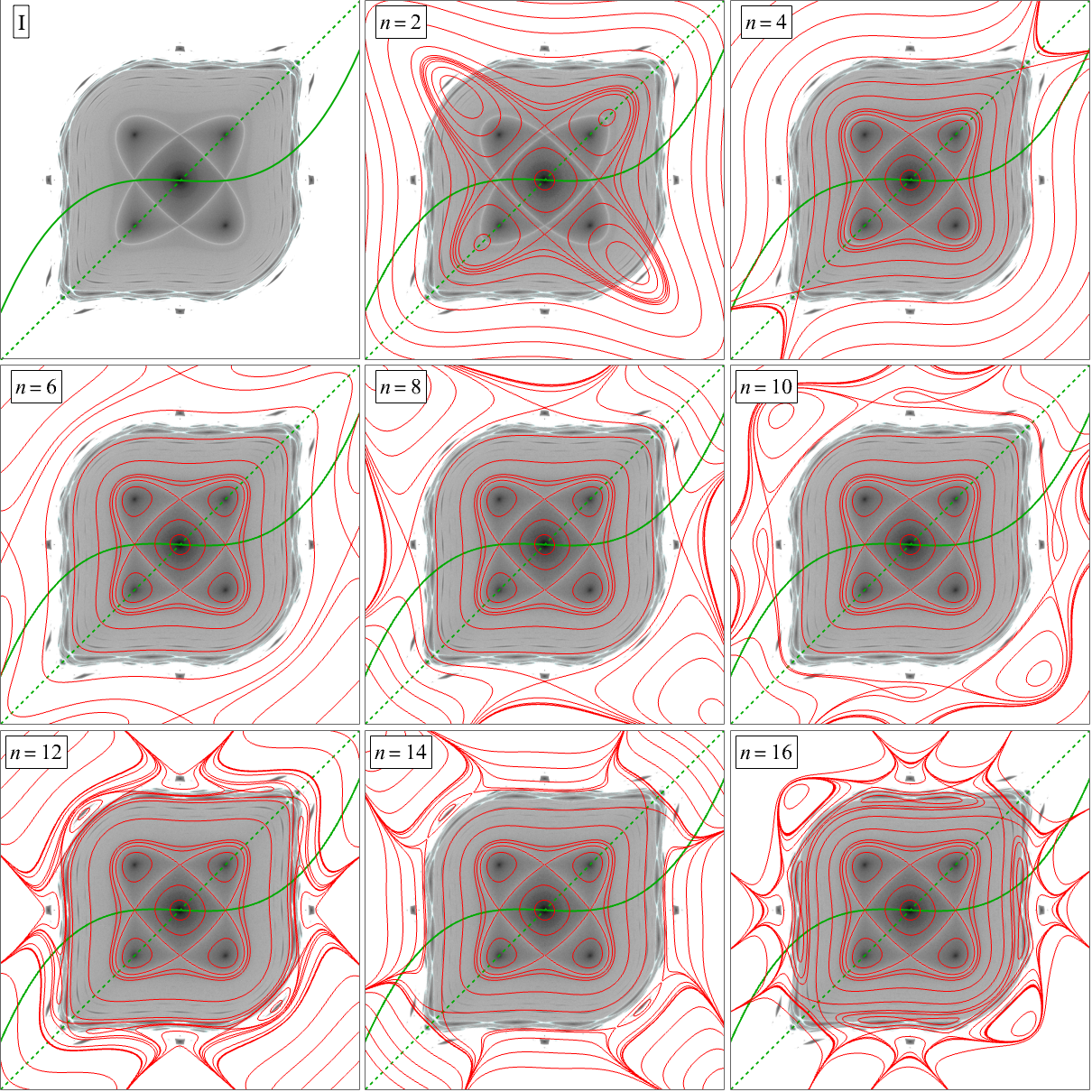}
    \caption{\label{fig:CaseI}
    Constant level sets of the averaged approximate invariant
    $\langle \mathcal{K}^{(n)} \rangle[p,q]$ (red), superimposed
    on the phase-space portrait from Fig.~\ref{fig:NuQExamples}.
    Shown for Case I: the cubic H\'enon map with $a = -1/10$.
    }
\end{figure*}
\begin{figure*}[h!]
    \centering
    \includegraphics[width=\linewidth]{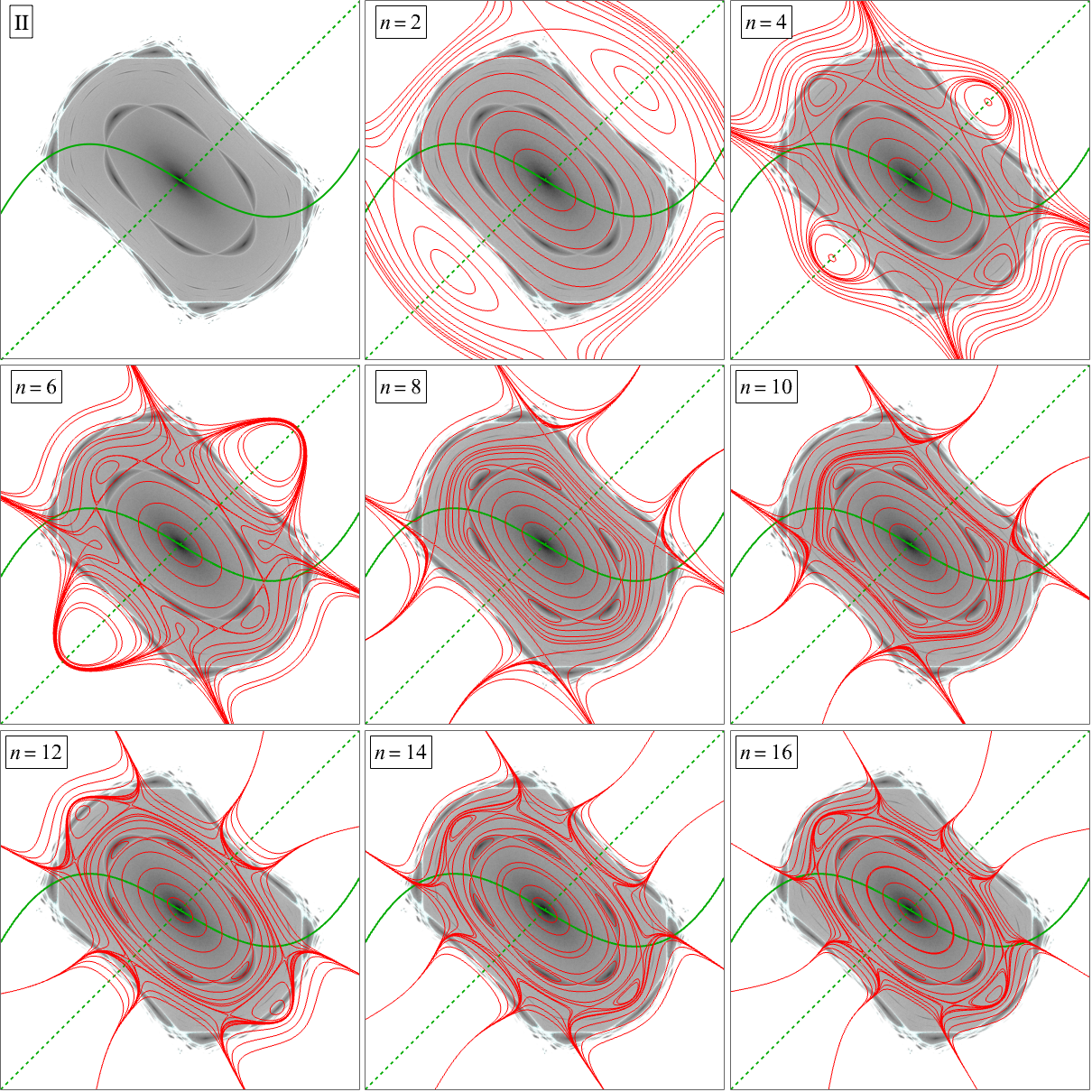}
    \caption{\label{fig:CaseII}
    Same as Fig.~\ref{fig:CaseI}, but for Case II: the cubic H\'enon
    map with $a = -6/5$.
    }
\end{figure*}

$\,$\newpage
$\,$\newpage

\begin{figure*}[t!]
    \centering
    \includegraphics[width=0.9\linewidth]{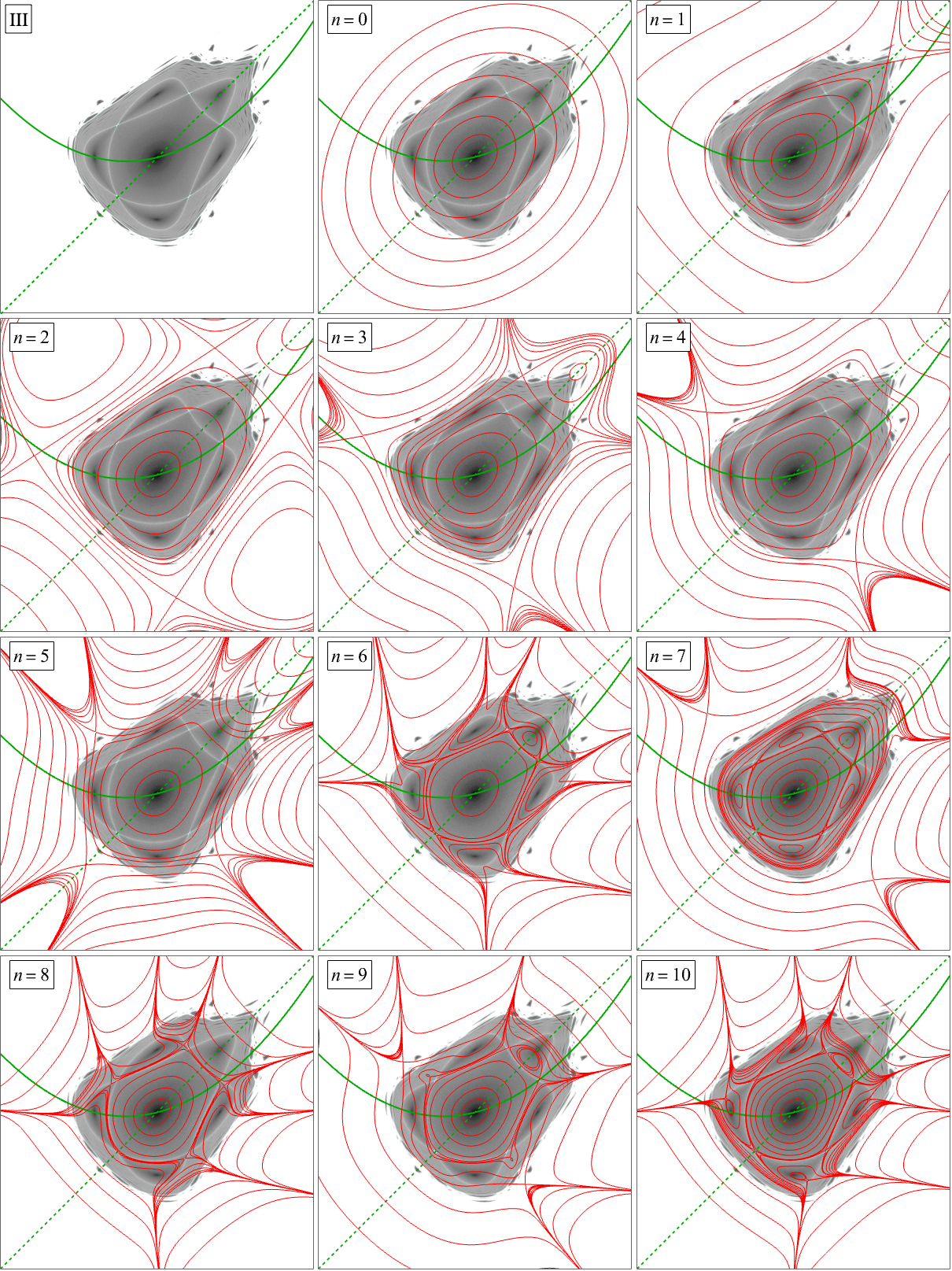}
    \caption{\label{fig:CaseIII}
    Same as Fig.~\ref{fig:CaseI}, but for Case III: the quadratic H\'enon
    map with $a = 1/2$.
    }
\end{figure*}

\section{\label{secAPP:SM}Approximate Invariants via the
Square Matrix Method}

\begin{figure*}[h!]
    \centering
    \includegraphics[width=\linewidth]{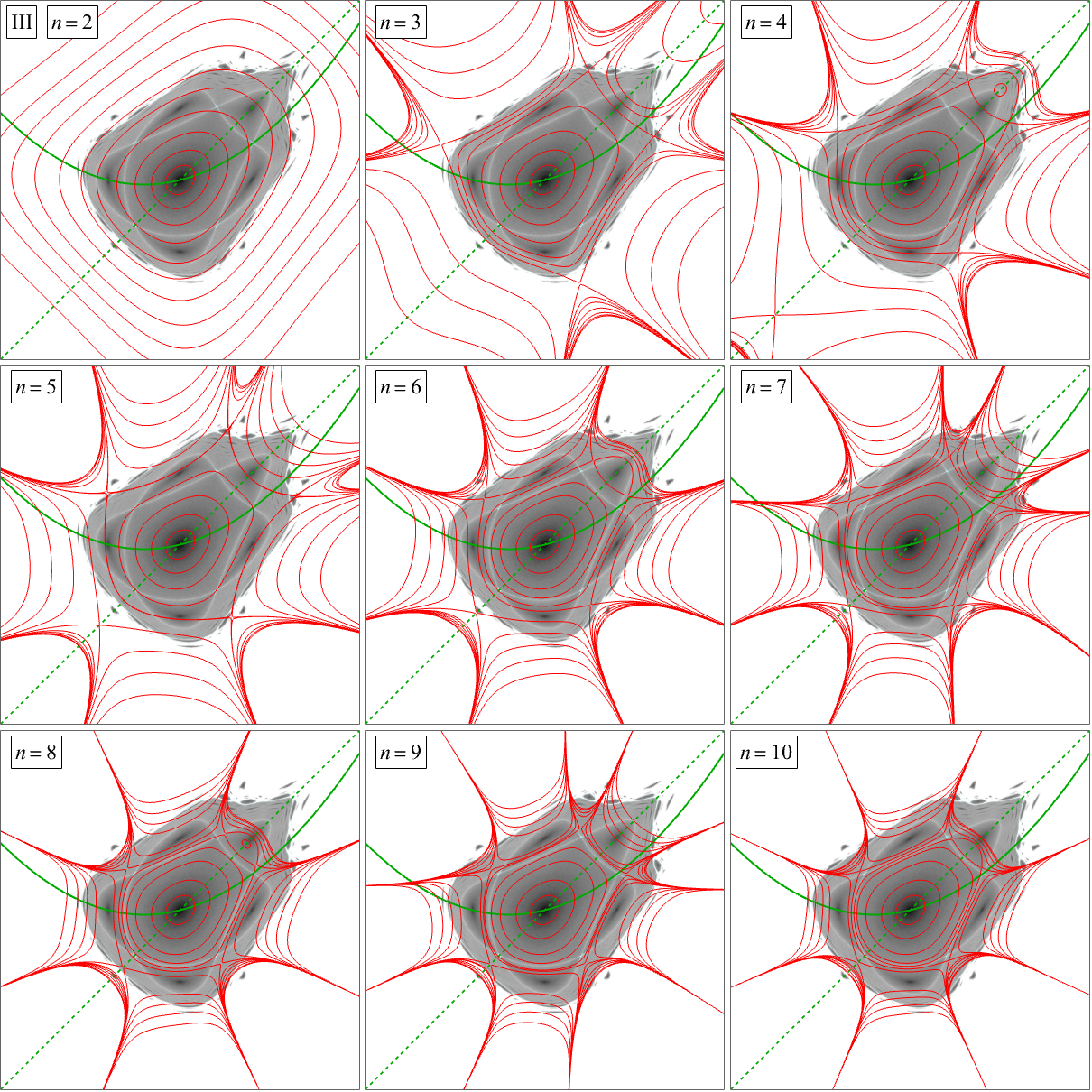}
    \caption{\label{fig:CaseIIIsm}
    Same as Fig.~\ref{fig:CaseIII}, but for the approximate
    invariants obtained via the square matrix method.
    }
\end{figure*}
\begin{figure*}[t!]
    \centering
    \includegraphics[width=0.9\linewidth]{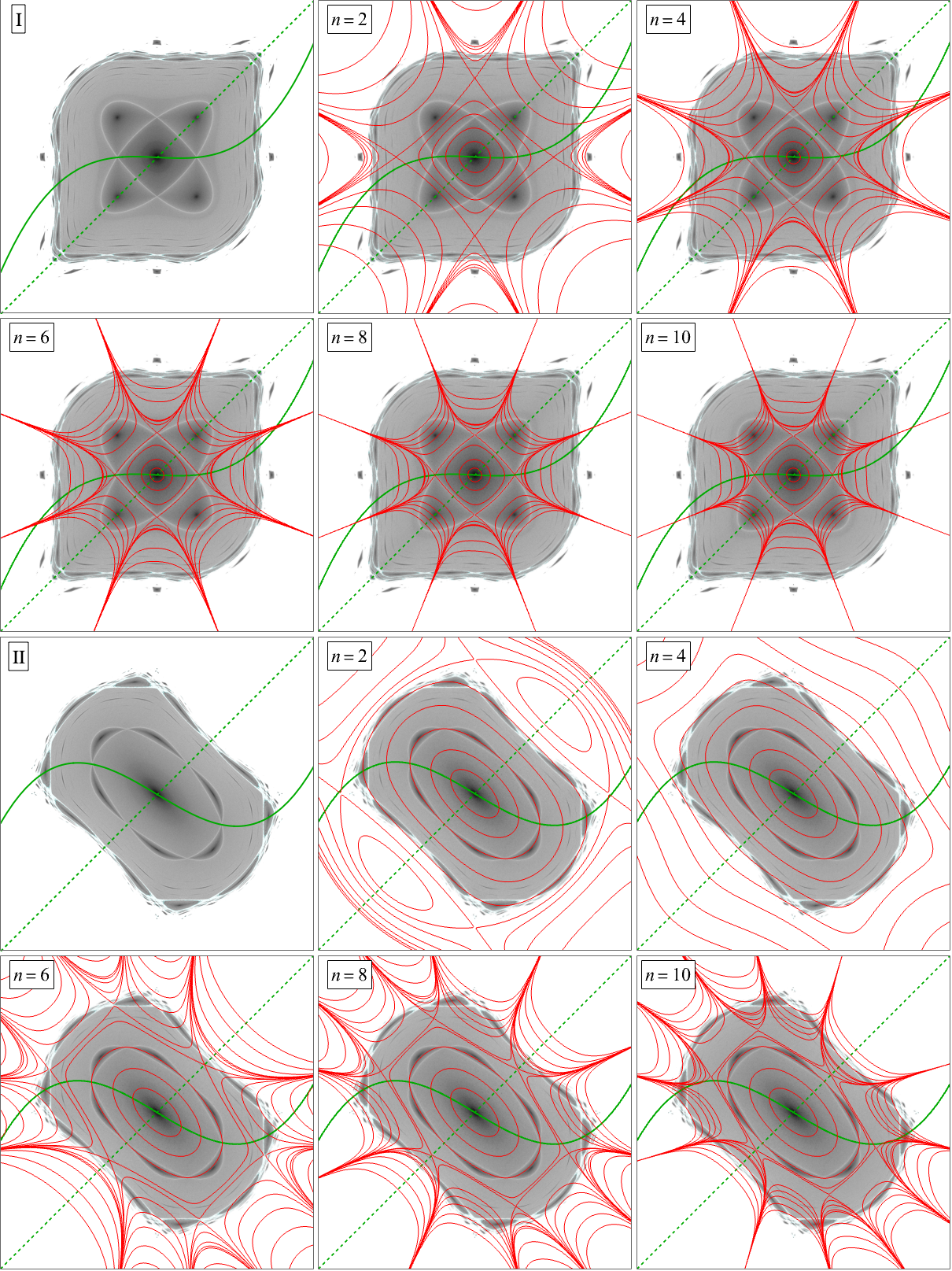}
    \caption{\label{fig:CaseI-IIsm}
    Same as Figs.~\ref{fig:CaseI} and \ref{fig:CaseII}, but for
    the approximate invariants obtained via the square matrix
    method.
    }
\end{figure*}



\newpage
%

\end{document}